# A Dynamic Multimedia User-Weight Classification Scheme for IEEE_802.11 WLANs

Ahmed Riadh Rebai[1] and Saïd Hanafi[2]

[1]Department of Electrical & Computer Engineering, Texas A&M University at Qatar, Doha, Qatar
`riadh.rebai@qatar.tamu.edu`
[2]LAMIH Laboratory, University of Valenciennes & Hainaut-Cambrésis, Lille, France
`said.hanafi@univ-valenciennes.fr`

*ABSTRACT*

*In this paper we expose a dynamic traffic-classification scheme to support multimedia applications such as voice and broadband video transmissions over IEEE 802.11 Wireless Local Area Networks (WLANs). Obviously, over a Wi-Fi link and to better serve these applications– which normally have strict bounded transmission delay or minimum link rate requirement – a service differentiation technique can be applied to the media traffic transmitted by the same mobile node using the well-known 802.11e Enhanced Distributed Channel Access (EDCA) protocol. However, the given EDCA mode does not offer user differentiation, which can be viewed as a deficiency in multi-access wireless networks. Accordingly, we propose a new inter-node priority access scheme for IEEE 802.11e networks which is compatible with the EDCA scheme. The proposed scheme joins a dynamic user-weight to each mobile station depending on its outgoing data, and therefore deploys inter-node priority for the channel access to complement the existing EDCA inter-frame priority. This provides efficient quality of service control across multiple users within the same coverage area of an access point. We provide performance evaluations to compare the proposed access model with the basic EDCA 802.11 MAC protocol mode to elucidate the quality improvement achieved for multimedia communication over 802.11 WLANs.*

## 1. INTRODUCTION

The IEEE 802.11 standard MAC mechanisms are considered to be insufficient for achieving a reasonable Quality of Service (QoS) under high traffic loads [1]. QoS has different meanings; from the user's awareness of the service to a set of network parameters needed to accomplish a particular service class. In fact, users need to obtain high end web services such as streaming video and audio even when they are moving around offices or wandering within a covered area. Future wireless access networks must be able to guarantee predefined levels of QoS, allowing the services mentioned above to be efficiently supported. Consequently, the real time multimedia applications provided by wireless connection requires a good amount of quality of services [2] support like; guaranteed bandwidth, delay, jitter and error rate. The standard 802.11protocol cannot deliver firm service guarantees while maintaining high network utilization, particularly under congested network conditions. As a result, QoS enhancements have been widely studied and evaluated. Currently, a QoS enhanced Medium Access Control (MAC) protocol standard, IEEE 802.11e, has been proposed for multimedia traffic demanding priority treatment mainly in terms of channel access delay [3]. The IEEE 802.11e version introduces a new significant infrastructure MAC mode: the Enhanced Distributed Channel Access (EDCA).

EDCA defines four priority classes or Access Categories (ACs) for transmitted data with different QoS requirements. Each AC uses a specific set of contention parameters, including the minimum and maximum Contention Window (CW), and the new Arbitrary Inter-Frame Space - IFS (AIFS), to compete the channel access. These backoff instances use different parameter settings to realize service differentiation. For example, a shorter CW is used for high-priority frames, so that, with high likelihood, they are transmitted before the low-priority ones. In the literature [4, 12, 13], performance evaluation of the EDCA scheme in WLANs has been reported through simulations. Specifically, in [13] the authors claim that under high loads of high-priority traffic, the EDCA scheme suffers from high collision rate and starves low-priority traffic. Moreover, recent studies have shown that the default values of the contention parameters are only good for scenarios with few high priority ACs and under moderate traffic loads, e.g. see [14, 16, 17]. The access point has the flexibility to adjust the contention parameters but no algorithm for this purpose has been addressed within the standard [3].

Therefore, EDCA can only provide probabilistic service assurances: high priority traffic should receive better service than low priority traffic. Since service classes are not strictly enforced, the servicing of low priority traffic can sometimes degrade the service designed to high priority traffic. Such occurrences are common, particularly under heavily congested conditions. This protocol estimates network conditions based on the aggregated network behavior of all traffic classes, reducing the effectiveness of differentiation techniques. Consequently, the main addressed problem in EDCA is how to adjust the contention parameters, depending on real network conditions, to achieve certain fairness between ACs and at the same time attain high channel utilization. Some works have proposed adaptive CW schemes, designed to coordinate MAC parameters between different stations. In Adaptive EDCA (AEDCA) [24], each node measures the collision rate in order to guide the adjustment of its CW. However, AEDCA still differentiates based on nodes rather than flows. A node's traffic can consist of multiple flows belonging to different traffic classes, with a large variation in traffic from one node to the next.

In this paper, we present a new access mechanism for IEEE 802.11e EDCA mode, based on an innovative inter-station priority to access the wireless medium. We verify the proposed model by comparing its performances with previous published results [4, 19, 22, 23]. The new 802.11e EDCA model includes the support of both packet priority and Mobile Station (MS) weight for data transmissions. Other than the AC resolution the new model allows the data packet sending based on the 'actual MS state' deduced from the recent admitted traffic in the network. Simulations were carried out to make a comparison of the QoS support provided by both IEEE802.11e standard EDCA and modified EDCA mechanisms. The first's theoretical results discern that proposed model outperforms the standard EDCA approach because of the impact of the inter-MS priority integration. In addition to the standard EDCA packet classification, the new adopted MS categorization will better manage the channel ability for an excellent QoS support. Thus, we are able to achieve a high performance WLAN transmission for packets with strict-QoS requirements. Furthermore, the proposed approach has a universal design and can be easily fitted with various MAC protocol modes to support strict-delay multimedia applications such as voice and video.

We organized the rest of this paper as follows; section 2 details the most recent and used channel access mode designed for IEEE 802.11 WLANs: the QoS-enhanced 802.11e EDCA extension [5, 6] and we point out its main inaccuracies and limitations. In section 3, we explain the proposed EDCA model by considering a new inter-station priority solution based on the previously adopted traffic. The performance evaluations of the new 802.11e EDCA mechanism are described in section 4 and compared with the literature ones [7, 18, 19, 23].

## 2. THE IEEE802.11E CHANNEL ACCESS PROTOCOL

This section explains the most typical and recent channel access scheme used by the 802.11e edition. It is divided into two subsections giving, respectively, details and drawbacks corresponding to this model.

### 2.1. Enhanced Distributed Channel Access (EDCA) Mode

In [5], several changes to the Enhanced Distributed Coordination Function (EDCF) protocol parameters and nomenclature were made: mainly the eight Traffic Categories (TC) in EDCF are now mapped onto four access categories in the Enhanced Distributed Channel Access protocol. In [8], eight traffic priorities are used which are specified in IEEE 802.1D [9]. In [3, 6], only four queues are used due to mapping the eight priorities to four ACs. As in Distributed Coordination Function (DCF), the backoff counter is stopped when the medium is sensed busy and is decremented if the medium is sensed idle for at least an AIFS period. The problem of repetitive collisions requires incrementing the contention window parameter value, $CW_{min}$, as in equation 1 while $CW_{min}$ and $CW_{max}$ depend on the AC traffic class.

$$CW_{min}^{new}[AC] = 2 \times CW_{min}^{old}[AC] + 1 \qquad (1)$$

EDCA sets up four queues per stations. Different ACs on one station with their own configuration parameters (AIFS and CW) are shown in Figure 1. The shown Voice, Video, Best Effort and Background categories are referenced, respectively, by AC_VO, AC_VI, AC_BE, and AC_BK. From now and to be simpler these ACs (AC_VO to AC_BK) will be numbered correspondingly from 0 to 3. Note that this may still lead to a collision if two ACs from the same node (user) are allowed to send at the same time. This collision is solved by a virtual scheduler that grants access to the AC with the highest priority.

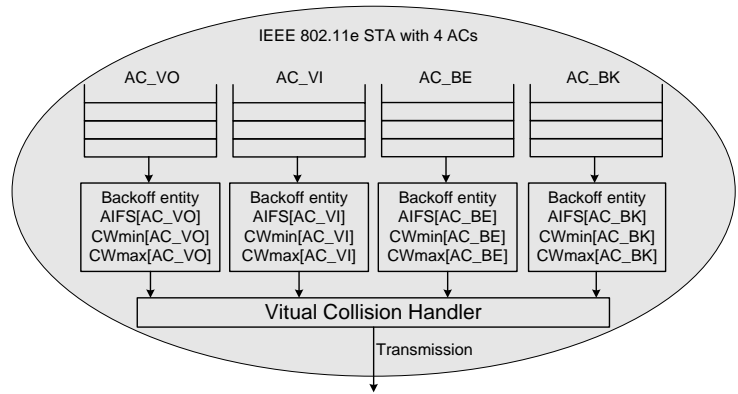

Figure 1. The EDCA channel access structure for the IEEE 802.11e MAC layer

The packet service differentiation is assured using an inversely proportional classification of $AIFS[i]$ as in equation 2.

$$AIFS[0] < AIFS[1] < AIFS[2] < AIFS[3] \qquad (2)$$

while $AIFS[0] > PIFS$ (PCF-mode IFS), with $CW_{min}[i]$ and $CW_{max}[i]$ for $i = \{0,..,3\}$, are classified as described in equation 2. Therefore, the EDCA channel access scheme depends on AC categories like in Figure 2.

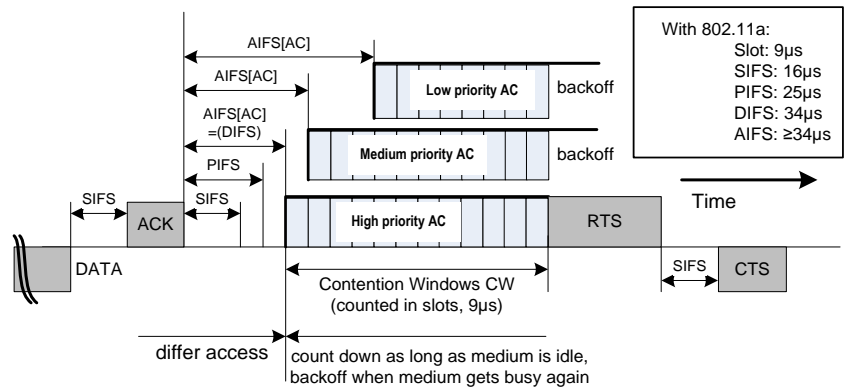

Figure 2. Inter-Frame Space relationships in IEEE 802.11e

## 2.2. Drawbacks of the 802.11e EDCA model

The typical approach for providing QoS in the IEEE 802.11e MAC, is to tune one of the three 802.11 MAC parameters, (i.e. Backoff Interval (BI), CW, and AIFS), based on certain fair queuing mechanisms. The BI and CW mechanisms are designed to control the backoff process by choosing an appropriate BI for different classes of traffic [10, 11]. The AIFS mechanism uses a different approach in which different IFS values are used to control the contention behavior of frames and no backoff process is involved. The DDRR algorithm [11] is based on Deficit Round Robin to translate user requirements into the IFS parameter of 802.11 MAC. So far, the standard EDCA model does not provide a proportional fairness service. Adaptive Fair Enhanced DCF (AFEDCF) [10] is the only mechanism that implements fairness among flows of the same priority for 802.11e EDCA. However, the AFEDCF procedure does not provide weighted fair service to different ACs and, therefore, cannot be considered as proportional fairness mechanism.

We point out that in EDCA, packet access qualifications are considered only for Intra-node traffic streams, and not provided for different user stations in the network. In other words, no distinction was taken into account by the standard mechanism between two active stations that are transmitting simultaneously data packets with same AC. Consequently, we can identify two major problems:

- Parallel AC Access: this describes the situation when packets from the same AC across multiple stations are transmitted concurrently within the same WLAN, resulting in possibly long contention delay (the ACs are using the same parameter values for the channel access tentative). This issue will increase the access conflict probability between transmitters having data with the same priority, and consequently, nullify most of the performance gains achieved by the EDCA technique.

- Static AIFS Assignment: To explain this deficiency in EDCA, we can easily observe that even when higher priority transmissions (ACs) are inactive, the lower priority packets will wait for a longer AIFS value before the backoff interval counts down to zero before they will be effectively transmitted in the network. As a result, the aggregate throughput degrades considerably when higher priority transmissions are missing. Additionally, the Backoff procedure for CW determination needs to be tuned according to the traffic load. This also introduces uncertainty in the performance of the MAC procedure.

International Journal of Computer Networks & Communications (IJCNC) Vol.3, No.2, March 2011

Various research studies [14-17] were performed to address the major problem linked to the IEEE_802.11 WLANS on how well the network can support quality of service (QoS). Specifically, these analyses investigate the network's performance in terms of maximum protocol capacity or throughput, delay, and packet loss rate. Although the performance of the 802.11 protocol, such as throughput or delay, has been extensively studied [14, 17] in the saturated case, it is demonstrated that maximum protocol capacity can only be achieved in the non-saturated case and is almost independent of the number of active nodes.

A deep review was directed in [15] on the most important problem in the IEEE 802.11 DCF-based WLAN. By analyzing packet delay, consisting of medium access control service time and waiting time, accurate estimates were derived for delay and delay variation when the throughput increases from zero to the maximum value. Furthermore, it was shown that the channel busyness ratio provides precise and robust information about the current network status, which can be utilized to facilitate QoS provisioning. Consequently, by controlling the total traffic rate, the original 802.11 protocol can support strict QoS requirements, such as those required by VoIP packets or streaming video, and at the same time achieve high channel utilization.

In [16] authors investigated the performance of model-based estimation over multi-access networks and emphasizes on the estimation Mean Squared Error (MSE) while using different channel access schemes: probabilistic (random access), deterministic (round-robin scheduling), and combined (grouped channel access). Estimation MSE, its asymptotic behavior and stability condition are derived for different channel access methods. The given quantitative discussion provides guidelines to design the communication logic for those control systems built on top of multi-access networks.

Many recent works were proposed [18-23] to outperform the QoS provisioning capacity of such networks. New features were added to satisfy the QoS requirement of established higher priority connections, while protecting the minimum reserved bandwidth of traffic flows with lower priority at the same time. In fact, the authors [19, 21] proposed an adaptive contention window adjustment mechanism which dynamically adjusts the maximum and minimum contention window size based upon the equivalent established connection number observed by each individual access category. Such approaches reduce the channel contention probability and packet access delay effectively. An enhanced EDCA scheme was presented in [20] which dynamically adjusts the two EDCA parameters: contention window and backoff time counter (BC), according to channel load conditions and QoS requirements of network services. This technique protects the standard backoff scheme from post-backoff, upon-collision and post-transmission phases respectively by differentiating real-time and non real-time frames. Specifically, frozen rate is calculated to estimate the channel load conditions during EDCA backoff process and then used to guide CW and BC adjustment.

Another suggested algorithm for EDCA [22] that, given the throughput and delay requirements of the stations that are present in the WLAN, computes the optimal configuration of the EDCA parameters. Then authors proposed a mechanism to derive the optimal configuration of the EDCA parameters, given a set of performance criteria for throughput and delay. In [23] an adaptive cross layer technique was proposed that optimally enhances the QoS of wireless video transmission in an IEEE 802.11e WLAN. The given optimization takes into account the unequal error protection characteristics of video streaming, the IEEE 802.11e EDCA parameters and the lossy nature of wireless channel.

In the following section, we propose new solutions contesting the above described problems, by introducing a new inter-node classification to make a better arrangement between traffic categories on the MAC-layer access process.

## 3. THE NEW INTER-NODE CLASSIFICATION MECHANISM

This section gives details of proposed techniques designed to deal with both problems discussed previously. The first subsection explains the design of an efficient channel access scheme for IEEE 802.11 networks. Then, we specify a second solution related to the multimedia traffic inactivity problem that can be integrated easily in the current data flow differentiation scheme.

### 3.1. The modified EDCA channel access technique for a finest QoS support

As discussed before the standard EDCA algorithm, has proven its capabilities for the service differentiation over the WiFi networks and has remedied the problems of virtual collisions (collisions between two different ACs in the same station). Consequently our approach is mainly based on it. To solve the first introduced problem (parallel transmissions of same-priority packets) we propose to add a (user) station ID as a new classification parameter for the transmitted packets. Thus, we introduce active station ID as another categorization, beside AC, to uniquely classify each transmission in the network. The new categorization scheme, as shown in Figure 3, will sort same service class flows for different simultaneous node transmissions. The new inter-node classification will reduce considerably the number of collisions that will be happened between the same AC access attempts. In the proposed method, we consider local information based on linked station traffic history. In other words, we add a prediction factor as decision part for MAC parameter values choice.

We count for all network active nodes, the number of transmitted packets $S_i$ for each $AC[i]$. This new parameter $S_i$ value will express the new station ID that will be designated by class $j$ in the rest of this paper. Each station will be categorized based on this new sorting parameter $j$ compared to the other stations accepting similar flow transmissions. Then, the station behavior and its channel access parameter values will change depending on the new inter-node classification $j$. Thus, the new $AIFS[i]$ value will not be fixed, rather, it will switch dynamically between $AIFS[i-1]$ and $AIFS[i+1]$ intervals depending on the $S_i$ counters. So, the $AIFS[i]$ value will be conditioned by the new station class computation. This new class $j$ provides the packet importance status relative to other same AC transmissions. There are two consequences of the proposed modification: First, equivalent AC flows will not have the same channel access priority due to the new AIFS configuration; Second, they will not be identically transmitted by different MSs. As a result, contention due to the parallel access will be reduced considerably for similar AC transmissions, and delays caused by collisions will be reduced substantially.

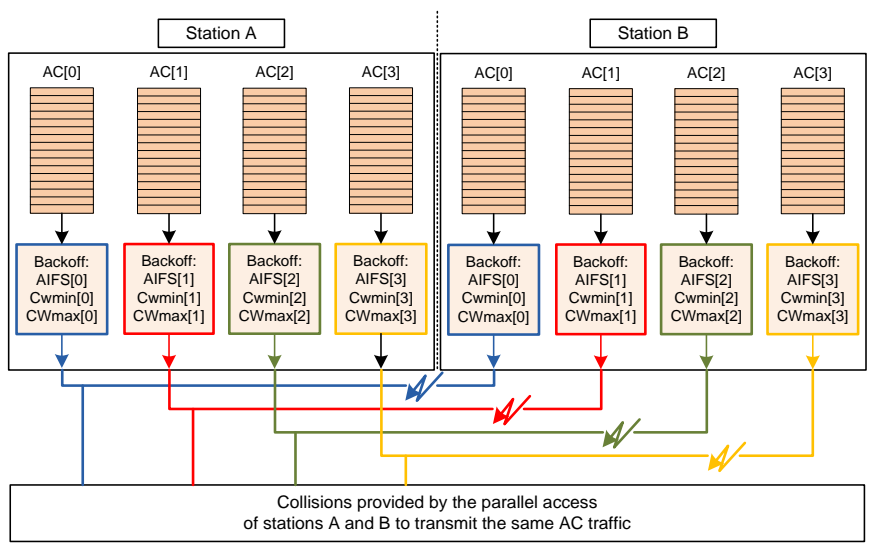

Figure 3. Collisions produced by the same AC access attempts in EDCA mode

The AIFS classification will not be proportional only to the traffic AC priority $i$, however it will depend also on the new station class $j$. The new *AIFS*[$i$][$j$] value will be more expressive in terms of traffic management and QoS handler. Also, it will maintain the flow classification ability as shown in equation 3 (lexicographical ordering).

$$AIFS[i][0]... < AIFS[i][j] < ... < AIFS[i][n-1] \quad (3)$$

with $AIFS[i-1][n-1] < AIFS[i][0]$, and $AIFS[i][n-1] < AIFS[i+1][0]$, where $i \in \{0,..., 3\}$ is the traffic category and $j \in \{0,..., n\text{-}1\}$ is the active station class.

We note that, the modified AIFS organization is still conserving the existing strict inter-traffic classification. In fact, it gives packets the priority to access channel depending firstly on their ACs. In addition, it prioritizes stations with frequently QoS transmissions than other nodes. *CW*[$i$] parameter values will not change by the new arrangement and will be always depending on traffic categories. The Figure 4 shows the new channel access scheme for the modified EDCA model.

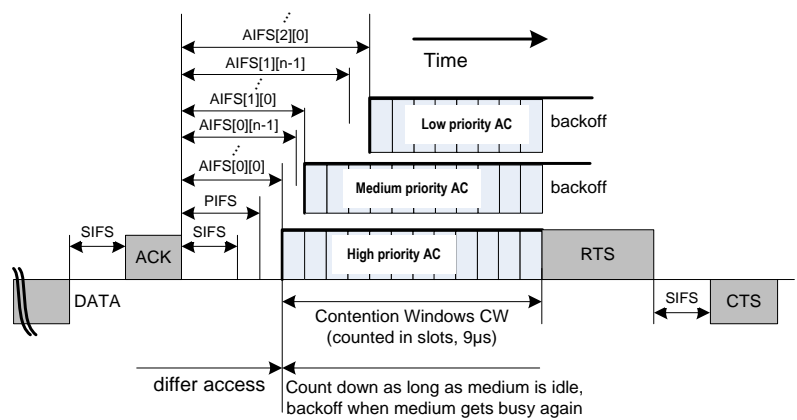

Figure 4. New Inter-Frame Space relationships in revisited EDCA mode

We specify the new station class $j$ computation method. It's fundamentally based on the local station counters $S_i$ examination and exploitation. The traffic counter $S_i$ settled for each *AC*[$i$] will reproduce the QoS trace in the recent traffic adopted by each station in the network. We remind that to conserve the distributed nature of the IEEE802.11 network, the new class $j$ computation must be calculated only from local station parameters. Consequently, we produce dynamic and significant decisions on the inter-station classification linked to CSMA 802.11 architecture constraints (without external network indications). The number of inter-station classes is set to $n$ ($0 \leq j \leq n\text{-}1$). Therefore, we establish $n$ new *AIFS*[$i$] values associated to each *AC*[$i$]. Accordingly, when an active station has a data packet with a flow priority *AC*[$i$] to be transmitted, it will proceed with its emission based on the new inter-station class $j$ associated to the actual station and the standard flow priority $i$ related to the data frame. As a result, the station will use the new *AIFS*[$i$][$j$] value for the channel access according to equation 3.

In practice, we note that the number of station classes, noted by $n$, cannot exceed 6 ($1 \leq n \leq 6$), because of the limitation of usable AIFS values [12, 13]. In fact, the modified EDCA mechanism limits the number of classes $n$ to 6 (in 802.11e EDCA medium access technique limits the AIFS values to fourteen only, i.e. AIFSN from 2 to 15). This value makes the new AIFS classification more beneficial when a large number of active stations is noticed on the network. In addition, we point out that when the value of $n$ is equal to 1, the new technique becomes identical to the standard EDCA algorithm and performs similar 802.11e medium access scheme.

In this paragraph, we give details on the station class's affectation based on the local counters $S_i$ linked to each flow $AC[i]$. In the proposed design, we make the mobile classification procedure totally dynamic and dependent on the network activity (last accepted transmissions). Thus, we implement an active algorithm that modifies and adapts the new station class $j$ based on packet counter value $S_i$ calculated for each access category flow; this arrangement will eliminate the case when all stations move to a higher or a lower priority class after few network transmissions, and so, the new mechanism will be identical to the standard procedure and will not have any advantage. Therefore, a mobile station can switch dynamically from class $j$ to a higher priority $j$-1 or a lower one $j$+1 depending on its recent accepted transmissions. Hence, in the proposed solution we have included a new inter-station priority as a station-importance/station-weigh factor to better manage the channel access between equivalent $AC[i]$ traffic. This add-on procedure is implemented separately on each mobile node and does not require changes on the actual CSMA/CA channel access mode assumed by the IEEE802.11e standard. In fact, it will choose in a distributed way (depending only on local counters) the next channel access parameter's values.

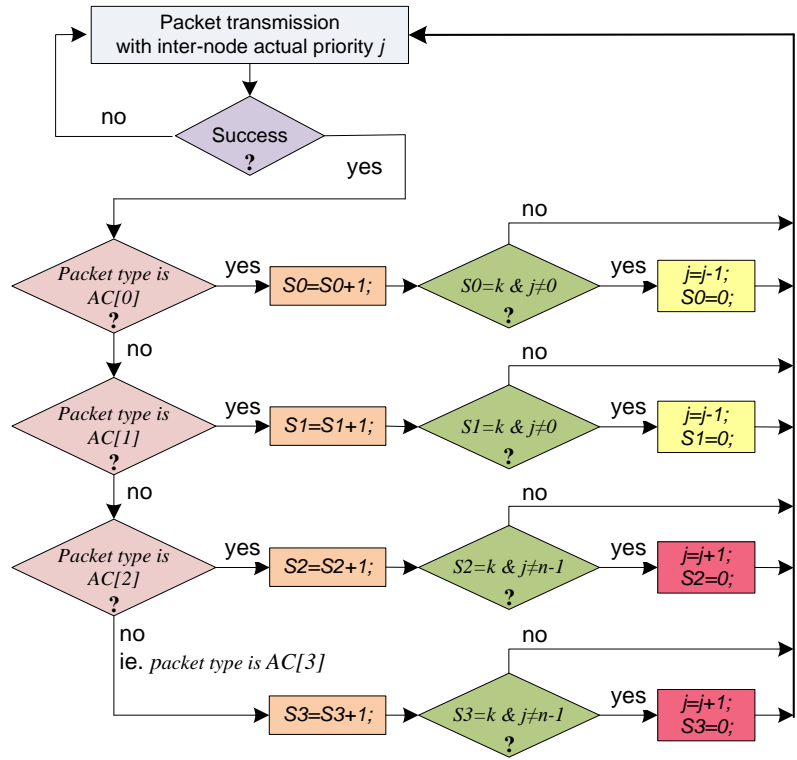

Figure 5. The new inter-node $j$ affectation in the revisited EDCA mode

The active inter-node priority affectation is detailed in Figure 5. Now, we give comments and procedure details linked to the modified EDCA method:

- The $k$ parameter is defined as the decision factor expressing the value limit for implemented local traffic counters $S_i$. Once it's reached by one of the traffic packet counters ($S_0$, $S_1$, $S_2$ or $S_3$), the mobile station switches from one inter-node class to another priority depending on the responsible traffic counter.

- If the $S_0$ or $S_1$ traffic counter reaches the value of $k$, we conclude a recent important multimedia traffic. Consequently, we allow the actual station more pertinence than other network units, and then, we increase its actual class $j$ to a higher $j$-1 priority.

- In the other hand, when the $S_2$ or $S_3$ counter reaches the $k$ value, the station commutes from $j$ to a lower $j$+1 class. Therefore, we conclude that next transmissions of the corresponding mobile are not significant and have not importance in the networks.

- When an inter-node class switching is occurred, only the $S_i$ flow counter responsible of this decision will be reset to zero.

- The initial class $j$ and the counter $S_i$ values are set, respectively, to a lower priority '$n$-1' and to 'zero' for each active mobile unit. Accordingly, mobile nodes will use the *AIFS*[$i$][$n$-1] value corresponding to their first transmission for *AC*[$i$] packets.

We point out that if the parameter $k$ value is chosen very small, the proposed process will be very predictive and the mobile station will be disturbed by a continuous class commutation, and vice versa. The $k$ and $n$ value's selection will be discussed in section 4.

## 3.2. Tolerant priority for low priority access categories

The problem linked to the high-priority traffic inactivity, as it was presented earlier in section 1, is less complex than the first one. In fact, when priority classes are not transmitted on the network, the lower ones still need to wait for the large AIFS value before the backoff interval will be counted down. Since the *AIFS*[$i$] parameter gives a strict priority between traffic categories, we decide to keep the same values, and so, maintain this classification. However, we choose to manipulate the *CW*[$i$] parameter value to minimize the waiting delay, and consequently, to maximize aggregate throughput. Then, we decide to provide the AC[0] and AC[1] backoff parameters a strict priority as specified by the standard. On the other hand, we produce a weighted fair service between the AC[2] and AC[3] classes by varying the corresponding CW backoff values as it is described in Table 1.

Our choice was agreed on the *CW*[$i$] value adjustment and not on the *AIFS*[$i$] parameter because of the nature of these factors. In fact, the second parameter has a fixed delay and relatively limited compared to the Backoff Interval that grows exponentially after each retransmission. Therefore, the EDCA channel access inefficiency caused by the strict priority will decrease considerably, while a classification between the low priority traffic is preserved. Furthermore, no modifications in the actual IEEE 802.11e algorithm are required, we record only a simple tuning procedure on the CW[2] and CW[3] parameter values.

Table 1. The EDCA parameters used in implementation

|            | AC[0] | AC[1] | AC[2] | AC[3] |
|------------|-------|-------|-------|-------|
| $CW_{min}$ | 3     | 7     | 15    | 15    |
| $CW_{max}$ | 7     | 15    | 1023  | 1023  |

## 4. PERFORMANCE EVALUATIONS

We firstly integrate the standard EDCA 802.11e algorithm as a C-code patch with the Network Simulator NS-2 platform [25]. Then, we elaborate the modified EDCA channel access as a second patch. We run each one of them using the same scripts to test and evaluate both implementations. The simulation scenario is an infrastructure Basic Service Set (BSS) mode, which consists of one base station and fifteen wireless nodes. The allowed data rate values used by MSs are in the range from 6 to 54Mbps as defined in 802.11a. Bidirectional streams of Voice and Video as well as Best effort and Background packets are transmitted between MSs using User Datagram Protocol (UPD) diffusion with Constant Bit Rate (CBR) traffic. The packet size is fixed to 1000Bytes. The simulation time and error model are chosen, respectively, to 10ms and Uniform Error Model with a grid of 300x300m$^2$. The selected propagation model is 'Two Ray Ground' and the *slot-time* unit is equal 6μs. We set the SIFS value to 8μs and the AIFS[0][0] parameter (corresponding to the most priority traffic in the network) to 20μs. During the simulations, the traffic transmission details adopted by MSs in are shown in Table 2. To better illustrate the new method and its improvements, the network transmissions will be based only on the proposed static data and not extra unexpected packets will be adopted by MSs during performed simulations.

Table 2. The Simulated traffic scenario

|  | MS1 | MS2 | MS3 | MS4 | MS5 |
|---|---|---|---|---|---|
| **Voice** | 20 | 30 | 60 | 10 | - |
| **Video** | 10 | 20 | - | 10 | 40 |
| **Best Effort** | 20 | - | 30 | - | 20 |
| **Background** | 40 | 100 | 20 | 90 | - |

We choose to simulate different traffic scenarios on the selected nodes to compare the new EDCA procedure versus the standard one like in [7]. The per-station channel access results for both techniques concerning voice, video and best-effort traffic are shown respectively in figures 6(a), 6(b) and 6(c). The reported curves correspond to the elected stations controlling the channel access by setting the packets enumeration on the x-axis. Our aim by showing this fist result is to illustrate the channel admission difference and the station behavior distinction between both simulated EDCA algorithms. The *k* and *n* parameter values are settled, respectively, to 10 and 6. As a second result, we will discuss later on the choice of these parameter values.

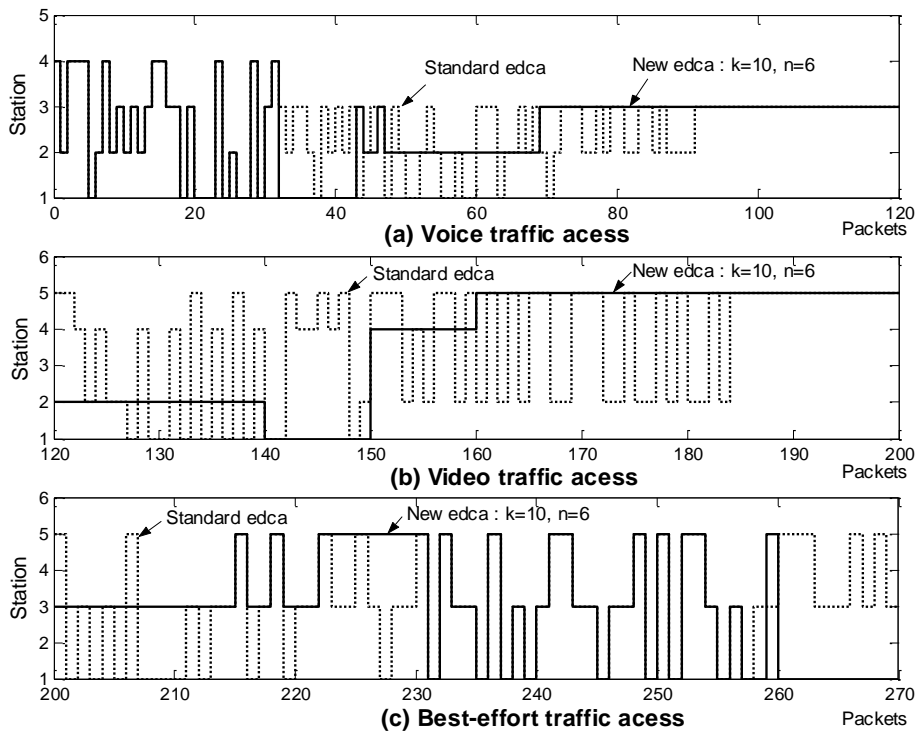

Figure 6. Traffic's channel admission

As exposed in figure 6(a), the voice traffic adopted on the channel using the new technique become dissimilar to the original one starting from the packet number 30. In fact, the proposed EDCA scheme promotes station 1, then station 2 and finally station 3 in order. This new deterministic channel admission will reduce considerably the number of collisions performed by the classic EDCA due to the use of identical access parameter values by the active stations. The dynamic inter-station classification is launched with the first transmitted packet. The MSs start switching between allowed classes $j$ depending on the adopted traffic, and so, the access parameter values will change concurrently. Thus, MSs will have different access priority for the same traffic category unlike the original technique. By examining the subplots 6(b) and 6(c), we figure out easily the impact of the new classification. Actually, the channel access tentative by active stations is prearranged depending on the nearest past of admitted transmissions on the network. For example, from packet number 223, station 3 reduces switches to a lower class by transmitting many Best-effort packets. Therefore, station 5 becomes more pertinent with a higher instantaneous class and gets an 'exclusive' channel access for next immediate transmissions in absence of unexpected high priority data (new Voice or Video frames admitted by any active station). Accordingly, the new method does not change the classic EDCA 802.11e channel access functioning. However, it acts and deals with the responsible access parameter values to better organize the channel admission between same traffic transmissions. As consequences, collisions are reduced significantly and the aggregate throughput is improved.

In the present paragraph, we will focus on the $k$ and $n$ parameter values discussion. In figures 7(a) and 7(b), we evaluate the influence of the new inter-node class number by means of two scenario configurations. In the first scenario we choose to implement six different classes ($n=6$). The second configuration uses only two inter-station priority layers. We decide to evaluate these values because they are significant. In fact the first case ($n=6$) represents the limit that we can implement due to the SIFSN values restriction. In favour of the second configuration we select a fluctuated value ($n=2$) to have a significant result. For both implementations the decision factor $k$ is set to 10. The corresponding Video and Best-effort packet admissions in the network are shown, respectively, in figures 7(a) and 7(b). We point out that the new priority introduces an excellent classification between stations that do not have the same adopted traffic history. Also, we deduce that the best number of inter-station classes to deploy in the network is proportional and strongly related to the number of active stations. Thus, we can conclude that conducted results using ($n=6$) are more suitable than the other case (i.e. $n=2$). Actually, we state that by adopting only two inter-station classes is equivalent to use eight different traffic categories and MSs are still competing the channel access. By increasing the number of classes to six, the new inter-station sorting becomes more considerable (equivalent to 24 different ACs) and the collision rate is minimised. As a result, and assuming that in most of cases WiFi networks admit at least 10 active MSs simultaneously (with a limit of 32) [7, 10, 12], setting $n=6$ will be the most appropriate value choice of inter-station classes for 802.11 networks.

We illustrate in figures 7(c) and 7(d) the decision factor $k$ manoeuvring ant its effects, respectively, on Voice and Best-effort flow admissions. We study two different value arrangements ($k=10$ and $k=20$) with a fixed inter-station class number ($n=6$). Firstly, we detect that the $k$ parameter value cannot be very small otherwise the proposed technique will be extremely predictive, and then, the channel admittance mechanism will be troubled by several periodic changes of responsible access parameter values. In the other hand, it shouldn't be very large, so that, the new mechanism would be able to follow the network transmission progressions and be capable to encourage the packet access of multimedia-marked stations. Based on these remarks, we point out that the first simulated configuration ($k=10$) has the most excellent results and is more efficient than the second formation ($k=20$). Assuming that both third and fourth ACs are more present in the network, it would be suitable to propose a different parameter value implementation in future work. We may apply a large decision factor value (e. g. $2*k$) for AC[2] and AC[3] categories instead of the $k$ value used for high priority ACs.

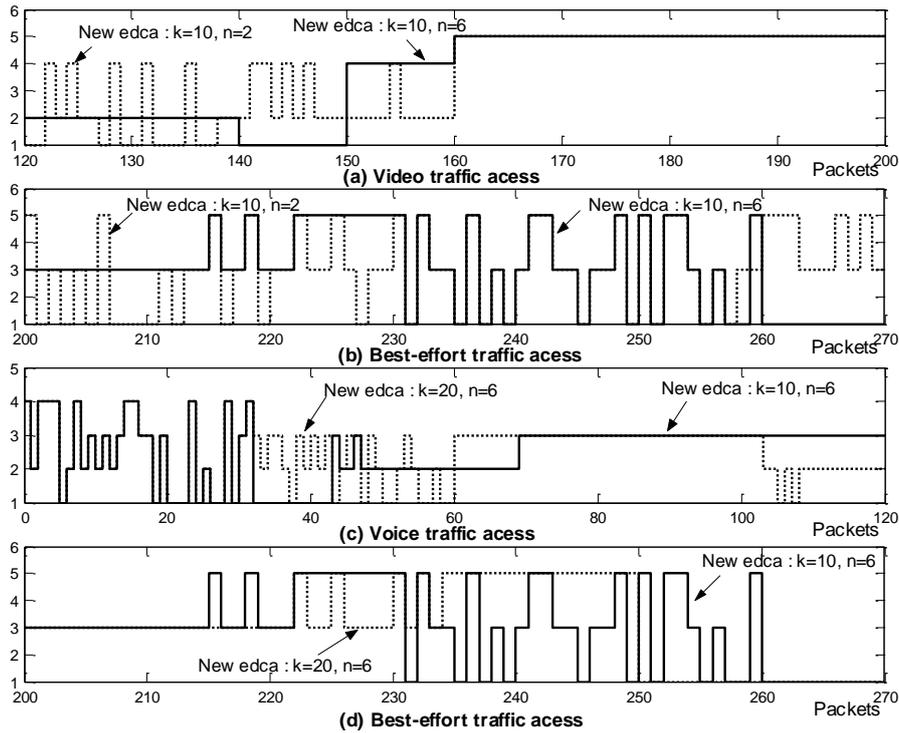

Figure 7. The *k* and *n* algorithm parameter discussions

A new network scenario, presented in Table 3, is simulated. Our aim is to illustrate the efficiency of the new model against the classic EDCA scheme for the high-priority traffic inactivity problem discussed in section 3.2.

Table 3. Network scenario without high priority Access Categories

|  | MS1 | MS2 | MS3 | MS4 | MS5 |
|---|---|---|---|---|---|
| **Voice** | - | - | - | - | - |
| **Video** | - | - | - | - | - |
| **Best Effort** | 200 | 600 | 300 | 500 | 200 |
| **Background** | 400 | 1000 | 200 | 900 | - |

In Table 4, we illustrate the channel occupation rate for the corresponding best effort and background flow transmissions (*k* and *n* are settled, respectively, to 10 and 6). The given results, presented in Table IV, show the ability of the new EDCA procedure to perform an excellent channel utilization compared to the classic EDCA technique. This obvious enhancement is achieved by the new fair priority tolerance solution. This resolution, as detailed in section III.B, deals with the CW parameter values to better support low priority data flows. Therefore, the standard EDCA algorithm is outperformed in terms of the channel use performance (e.g. a network exploitation gain more than 7% for Best-effort packets and around 3% regarding the Background traffic). Accordingly, the obtained rates reflect a significant time reduction for the large waiting delays achieved by MSs before start their low priority packet transmissions. As a consequence, the proposed scheme improves the network management when multimedia communications/diffusions are active (by adding a dynamic user-weight to each MS) or inactive (by inserting a weighted fair service for low-importance data flows). Also, for both simulation scenarios presented in Tables 2 and 3, we have calculated the overall transmission time. The total time achieved by the standard EDCA technique in the first and the second configuration is minimized, respectively, by 23.2% and 17.9% using the proposed scheme.

Table 4. Channel utilization rates for low priority traffics

|  | **Basic EDCA** | **New EDCA** *(k=10, n=6)* |
|---|---|---|
| **Best Effort** | 17.6% | 25.4% |
| **Background** | 12% | 14.9% |

To better illustrate one of the main advantages of the proposed algorithm, we drive new simulation series. In Figure 8, we summarize the collision rate completed on each AC during the data transmission. The network configurations implement the classic EDCA mechanism and the revised scheme in many versions: ($k=10$; $n=2$), ($k=20$; $n=6$), and ($k=10$; $n=6$). As shown in Figure 8, the collision rate counted for each different AC is reduced by all versions of the new customized EDCA model. The best values are reached by setting $k=10$ and $n=6$. This reduction contributes subsequently to the transmission latency minimization, and therefore, to the overall throughput diminution. Finally, we conclude that the modified technique includes both strict inter-node classification and weighted fair service solutions for, respectively, high and low priority access categories to order the channel admission between MSs, to minimize the waiting delay, and so, to maximize the aggregate throughput.

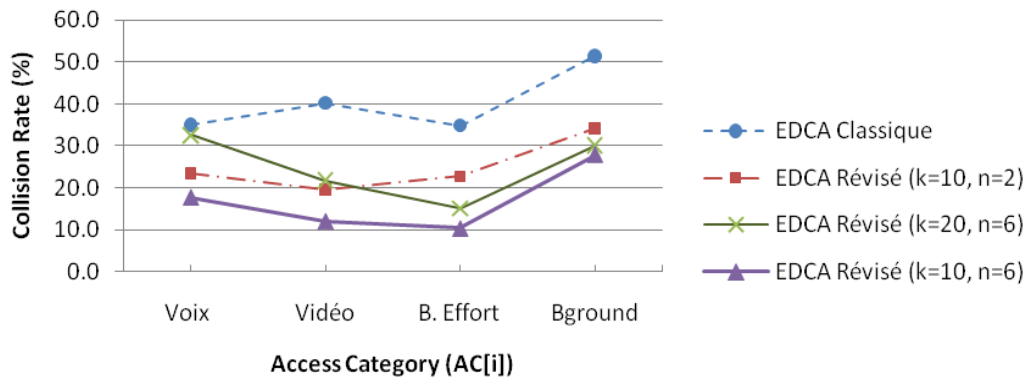

Figure 8. Collision occurrences on each Access Category

## 5. CONCLUSIONS

The aim of this paper is to enhance the support of multimedia applications such as voice and broadband video transmissions, which normally have a strict bounded transmission delay, and so, to improve the Quality of Service (QoS) scheme in IEEE 802.11e WLANs. Firstly, we have included a new inter-station classification to the EDCA channel access model, based on a modified inter-frame space arrangement. This characterization will produce a different multimedia rank for each mobile node depending on its precedent accepted traffic. As well, we focused on the multimedia traffic inactivity problem that produces a very large waiting delay for the background packet transmissions, and consequently, affects the channel utilization rate. We have provided a new tolerant fair service QoS mechanism for low priority data flows based on the backoff interval manipulation. We detailed the proposed algorithm and compared its performances with the classic EDCA 802.11e mechanism. The simulation results show that the revised mechanism outperforms the basic EDCA in terms of providing support to both strict priority and weighted fair service for, respectively, multimedia and low priority packets. Compared to other solutions proposed in literature, the revised EDCA model is easier to implement in real systems and has a better aggregate throughput. More importantly, the proposed mechanism adds a user-weight categorization between active stations, rendering it a good candidate to provide an excellent multimedia service and a QoS management for IEEE 802.11 WLANs.

**Authors**

**Ahmed Riadh Rebai** received his Ph.D. and M.Sc degrees in Computer and Network Engineering, respectively, from the University of Valenciennes - France, and the National engineering school of Sfax - Tunisia. He joined the Electrical and Computer Engineering Program of the Texas A&M University since January 2009. Previously, he held permanent teaching positions with University of Sciences and Polytechnic High School of Tunis - Tunisia. His current research interests are in network protocols design, network coding, vehicular communications, and quality-of-service (QoS) mechanisms in wireless networks.

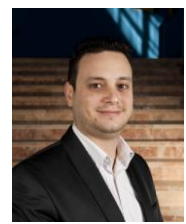

**Saïd Hanafi** holds a Full Professor position in Computing Science at Institute of Techniques and Sciences, University of Valenciennes and is currently in charge of the team Operations Research and Decision Support. His research lies in the design of effective heuristic and meta-heuristic algorithms for solving large-scale combinatorial search problems. His is interested in theoretical as well as algorithmic modelling and application aspects of integer programming and combinatorial optimisation and has published over 30 articles on the topic. His current interests revolve around the integration of tools from hybrid methods mixing exact and heuristics for solving hard problems.

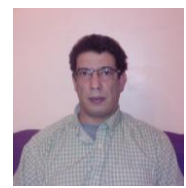